\documentclass[preprint,12pt]{elsarticle}
\usepackage{amssymb}
\newcommand{\refig}[1] {Fig.~\ref{#1}}

\journal{Journal of Terramechanics}
\begin{document}
\begin{frontmatter}


\title{Extending Granular Resistive Force Theory to Cohesive Powder-scale Media}
% Alternative: Extending Granular Resistive Force Theory to Cohesive Regime


\author[inst1]{Deniz Kerimoglu\fnref{label2}}
\author[inst2]{Eloise Marteau}
\author[inst1]{Daniel Soto}
\author[inst1]{Daniel I. Goldman}

\affiliation[inst1]{organization={School of Physics, Georgia Institute of Technology},
            addressline={837 State St.}, 
            city={Atlanta},
            postcode={30318}, 
            state={GA},
            country={USA}}

\affiliation[inst2]{organization={Jet Propulsion Laboratory, California Institute of Technology},
            addressline={4800 Oak Grove Drive}, 
            city={Pasadena},
            postcode={91109}, 
            state={CA},
            country={USA}}            
\fntext[label2]{Corresponding Author: dkerimoglu6@gatech.edu}

\begin{abstract}
Intrusions into granular media are common in natural and engineered settings (e.g. during animal locomotion and planetary landings). While intrusion of complex shapes in dry non-cohesive granular materials is well studied, less is known about intrusion in cohesive powders. Granular resistive force theory (RFT)—a reduced-order frictional fluid model— quantitatively predicts intrusion forces in dry, non-cohesive granular media by assuming a linear superposition of angularly dependent elemental stresses acting on arbitrarily shaped intruders. Here we extend RFT's applicability to cohesive dry powders, enabling quantitative modeling of forces on complex shapes during intrusion. To do so, we first conduct intrusion experiments into dry cornstarch powder to create stress functions. These stresses are similar to non-cohesive media; however, we observe relatively higher resistance to horizontal intrusions in cohesive powder compared to non-cohesive media. We use the model to identify geometries that enhance resistance to intrusion in such materials, aiming to minimize sinkage. Our calculations, supported by experimental verification, suggest that a flat surface generates the largest stress across various intrusion angles while a curved surface exhibits the largest resistance for vertical intrusion. Our model can thus facilitate optimizing design and movement strategies for robotic platforms (e.g. extraterrestrial landers) operating in such environments.\end{abstract}



\end{frontmatter}
\section{Introduction}
\subsection{Resistive Force Theory}
\label{sec:intro}
Interactions of solid objects with granular media occur in a wide range of contexts, from engineered settings to natural interactions. Granular media exhibit complex behaviors when interacting with a rigid body, driven by friction-dominated inter-particle forces. Such interactions produce complex flow and force responses, with the material displaying both solid-like and fluid-like properties. The principles governing the interaction of rigid objects moving within plastically deforming media have been applied in robotics to develop machines \cite{bhushan2009biomimetics,maladen2009undulatory,karsai2022real} and vehicles \cite{bekker1960off} capable of traversing granular surfaces. These complex interactions are fundamentally influenced by how local granular material properties generate global resistive forces on intruding objects.

To predict the role of granular terrain under rigid intrusions, one approach is to simplify the behavior of granular media into tractable force models. Granular resistive force theory (RFT), a reduced order frictional fluid model, has been proposed to provide quantitative predictions of quasi-static intrusion forces in dry, non-cohesive granular materials \cite{li2013terradynamics,maladen2011mechanical}. Despite the complex properties of granular materials, RFT stands out for its remarkable simplicity and effectiveness \cite{schofield1968critical,kamrin2010nonlinear,kamrin2012nonlocal,henann2013predictive}. RFT has been successfully applied to understand and improve the granular locomotion of wheeled systems \cite{agarwal2021surprising} and legged robots \cite{lee2020learning,chong2021coordination,treers2022mole}. 

RFT operates by linear superposition of the elemental forces obtained by discretizing the surface of the arbitrarily shaped intruders. The total resistive force is approximated as the sum of the resistive forces acting on each discretized element as if each element were independently moving. The physics underlying the success of granular RFT in predicting granular resistive forces has been numerically investigated in \cite{askari2016intrusion} through a frictional plasticity continuum PDE model. The continuum model quantitatively matched experimental intrusion data and RFT predictions using two key mechanical properties of dry granular media: a frictional yield criterion and the absence of cohesion between particles. Despite its success, RFT has only been applied to predict intrusion behaviors in non-cohesive materials. The generality and applicability of the continuum model have yet to be verified for dry cohesive substrates.

\begin{figure}[!ht]
    \centering
    \includegraphics[width=.9\columnwidth]{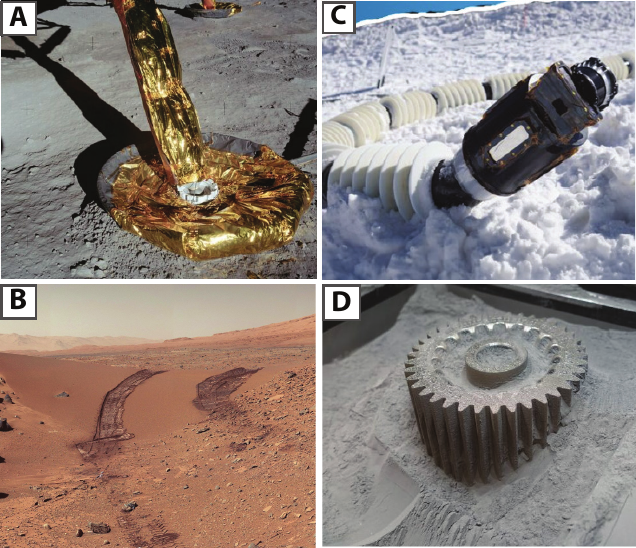}
    \caption{Examples of dry cohesive powder-scale media A) Planetary lander footpad on the moon surface, (B) Mars regolith (C) EELS snake robot on loose snow, and (D) Metal powder 3D printing. Photo credits: A) Apollo 11, Lunar Module Footpad B) https://en.wikipedia.org/wiki C) \cite{vaquero2024eels} D) https://www.heraeus-group.com/en/}
    \label{fig:intro_fig}
\end{figure}

Dry cohesive, very fine particulate media (particle diameter $~10-50~\mu m$) is common in natural and artificial environments. Examples include extraterrestrial settings with legged (\refig{fig:intro_fig}A) and wheeled (\refig{fig:intro_fig}B) systems on regolith, robotic systems navigating loose snow (\refig{fig:intro_fig}C), and metal powder in 3D printing processes (\refig{fig:intro_fig}D). Although the particles themselves are non-cohesive, the cohesion observed between dry powder particles arises primarily from Van der Waals forces or electrostatic interactions \cite{castellanos2005relationship,feng2003relative}. However, these forces can still be generated locally based on the rigid body's shape during intrusion. In other words, as predicted by \cite{askari2016intrusion}, when moving through cohesive powder the elemental forces on an intruder's segments are expected to depend only on its position, orientation, and direction of motion, without cross-correlation between segments. We hypothesize that the linear superposition principle of non-cohesive Resistive Force Theory (RFT) can also be extended to cohesive powders.

\subsection{Planetary Lander Footpads}
In the solar system, various fine regolith materials are known to exhibit cohesive properties. Geotechnical measurements and sample analyses from the Apollo missions have demonstrated that the lunar regolith exhibits considerable cohesion, particularly within its finer particles \cite{heiken1991lunar,slyuta2014physical}. On Mars, observations of trenches and scrapes created by lander scoops, footpad imprints, and rover tracks have revealed that Martian regolith exhibits cohesive properties \cite{moore1989viking,moore1987physical,team1997characterization,shaw2009phoenix,arvidson2014terrain,golombek2020geology,marteau2023soil}. In the outer solar system, Enceladus and potentially Europa eject materials from their subsurface oceans into their exospheres, with some of this material settling back onto the surfaces of these Ocean Worlds. Plume deposits are expected to consist of fine-grained ice particles that evolve via sintering, possibly transforming initially unconsolidated deposits into consolidated porous ice \cite{choukroun2020strength,molaro2019microstructural,dhaouadi2022discrete}. Future exploration of these icy moons by a lander \cite{mackenzie2021enceladus,hand2022science} will require understanding the interactions between their surface material and landing footpads during a touchdown event. This is particularly relevant to the South Pole of Enceladus, which is of great scientific interest, where freshly deposited plumes are loosely consolidated, making successful landings challenging. Few studies have been conducted to understand the material properties \cite{jabaud2024cohesive} and the interaction of landing gear with such weak and cohesive surfaces \cite{harmon2023predicting}. As robotic space exploration advances, the cohesive properties of these various regolith types have significant implications for mission formulation, influencing mobility, instrumentation, and the design of landing systems across diverse planetary surfaces. Terramechanics models used in planetary exploration have predominantly operated under the assumption that regolith is cohesionless, neglecting the cohesive properties present in extraterrestrial soils. Similarly, analyses of landing footpads have relied on outdated and conservative assumptions, primarily rooted in testing conducted during the Apollo era in the 1960s \cite{winters1968lunar}. Since then, there have been no comprehensive testing campaigns to evaluate landing footpad interactions with cohesive or complex regolith, highlighting a significant gap in our current understanding and design approaches for future planetary landers.

In this paper, we conducted systematic Resistive Force Theory (RFT) experiments in dry cohesive powders, measuring intrusion forces to demonstrate that these forces can be linearly superimposed.  We observe an approximately linear stress-depth relationship which we used to create stress per unit depth functions. We compared the stress profiles of cohesive and non-cohesive media and experimentally validated our powder RFT (pRFT) model. We observed that while the stress patterns in the cohesive powder were similar to those in non-cohesive granular media, the cohesive powder displayed significantly higher resistance to horizontal intrusions. To utilize the predictive capabilities of our model we identified geometries that enhance intrusion resistance in weak and cohesive surfaces, aiming to identify shapes for potential landers. Focusing on planetary lander footpad geometries we aimed to minimize vertical sinkage during a landing event on weak surfaces, which pose landing stability challenges. We first created various model footpad shapes and conducted extensive pRFT calculations with them. We discovered that a flat geometry, reflecting a disk-shaped footpad, generates larger stresses (force per unit area) than curved ones across a wide range of intrusion scenarios. Conversely, a curved geometry with an increased surface area generated more resistive force than a flat one, particularly under more vertical intrusion conditions. We validated our model's capabilities by experimentally testing diverse footpad geometries, observing a strong correlation between the experimental results and pRFT predictions. 

\section{Materials and Methods}
\begin{figure}
    \centering
    \includegraphics[width=1\columnwidth]{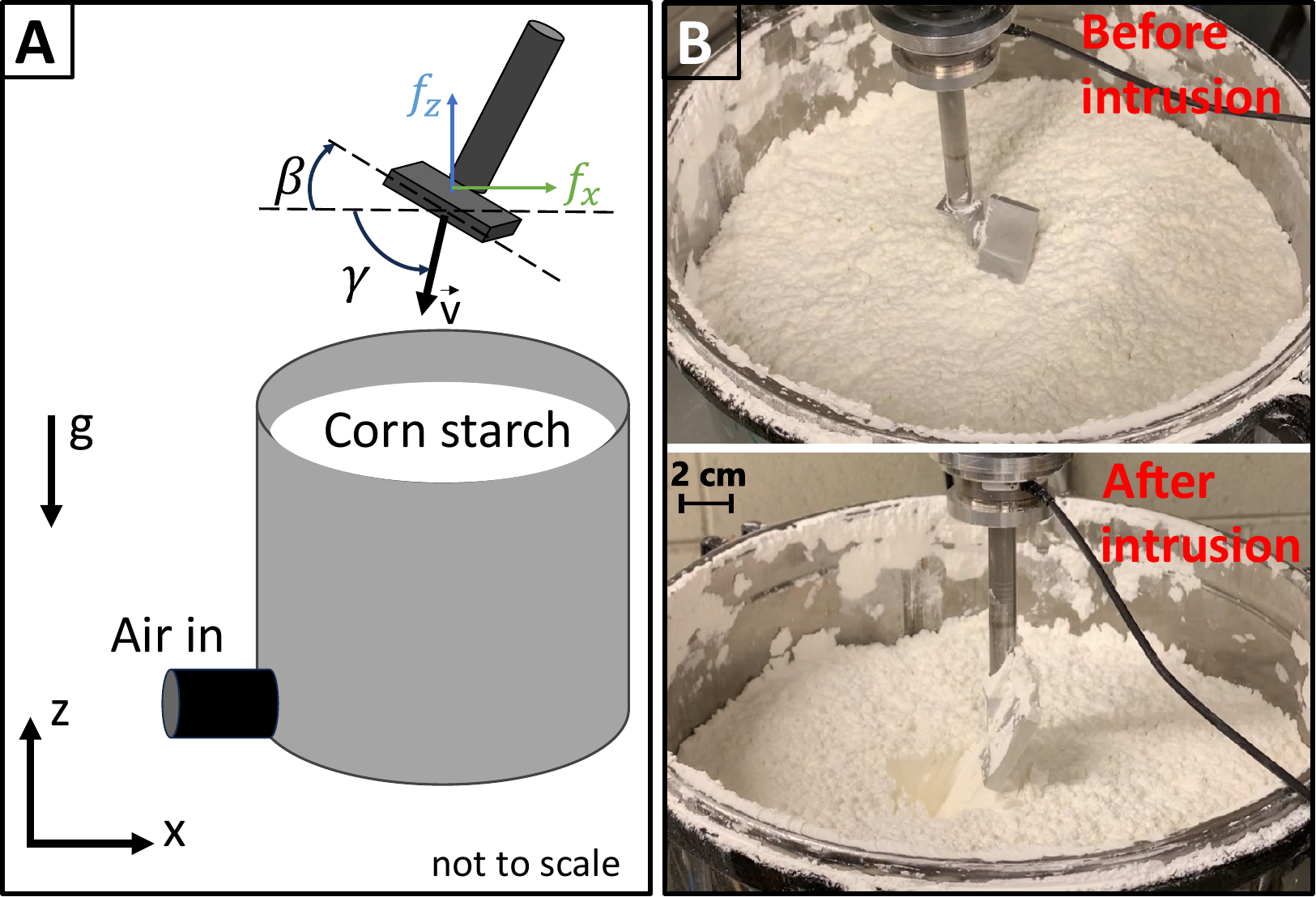}
    \caption{Apparatus for controlled intrusion experiments into dry powder. A) Diagram of the test platform used for powder intrusion experiments. The chamber, filled with cornstarch powder, is connected to an air blower that facilitates the creation of a loosely consolidated powder state. A robot arm, along with the support rod holding the plate element, controlled the plate's attack and intrusion angles. B) An image of the cornstarch-filled chamber before and after the intrusion of the rod attached to the aluminum plate. Notice the crater formed after the intrusion. The substrate around the intrusion site did not substantially reconfigure, retaining its post-intrusion configuration due to inter-particle cohesion even after the intruder was removed.} 
    \label{fig:apparatus}
\end{figure}

We developed an apparatus to perform systematic intrusion experiments \refig{fig:apparatus}A) into loosely consolidated dry powder. A stainless steel pressure chamber (30 cm in diameter and 40 cm in height) as shown in \refig{fig:apparatus}B is attached to an air blower. The chamber's interior features a honeycomb structure and a porous plastic layer, which facilitates the creation of a uniform airflow when the air blower is activated. We used cornstarch powder as a model dry and cohesive powder substrate and filled the chamber to a depth of 17 cm. Before each experiment, we carefully applied air fluidization and mechanical agitation to set the powder to a loosely packed state \cite{skonieczny2019rapid}. We measured the bulk volume fraction $\phi=0.35$ of the substrate by calculating the ratio of solid volume to occupied volume using the density of cornstarch particles ($1.34~g/mL$) \cite{fuentes2019fractionation}.
The intrusion experiments are conducted using a robot arm (Denso Robotics) moving at a constant speed, $v=1~cm/s$. We mounted a 6-axis force/torque sensor (ATI Industries Mini-40, SI-80) to the robot arm's end effector to capture powder resistive forces. A supporting rod with a connector is mounted on the force sensor, facilitating the attachment of an aluminum plate to the rod. The robot arm motion and the support rod connection enabled us to adjust the angle of intrusion $\gamma$ and angle of attack $\beta$ of the plate (dimensions 2.5 cm × 3.8 cm × 0.6 cm) as depicted in \refig{fig:apparatus}A). We measured the forces in the horizontal (x) and vertical (z) axes at $1$~kHz during plate intrusion experiments. We first measured the forces on the supporting rod as it moved through the powder without the plate attached for each $\gamma$ and $\beta$ combination. Next, we repeated the intrusion with the rod and plate, subtracted the rod-only forces, and divided the result by the plate's surface area to calculate the stresses applied only on the plate. We used the stresses in the region away from the surface and the bottom of the chamber. We performed three experiments for each combination of $\gamma$ and $\beta$ and used their means to obtain stresses $\sigma_{z,x}$ for $-\pi/2<\beta<\pi/2$ and $0<\gamma<\pi/2$. The positive range for $\gamma$ is selected because we are focused on the intrusion of objects. The stresses $\sigma_{z,x}$ for horizontal movements ($\gamma = 0$) were determined similar to \cite{li2013terradynamics}, by fitting the stresses to the average values in the steady-state regions at depths of $z$ = 3 cm, 6 cm, and 9 cm. 

\section{Results}
\subsection{Developing Powder Resistive Force Theory}
\begin{figure}[!ht]
    \centering
    \includegraphics[width=1\columnwidth]{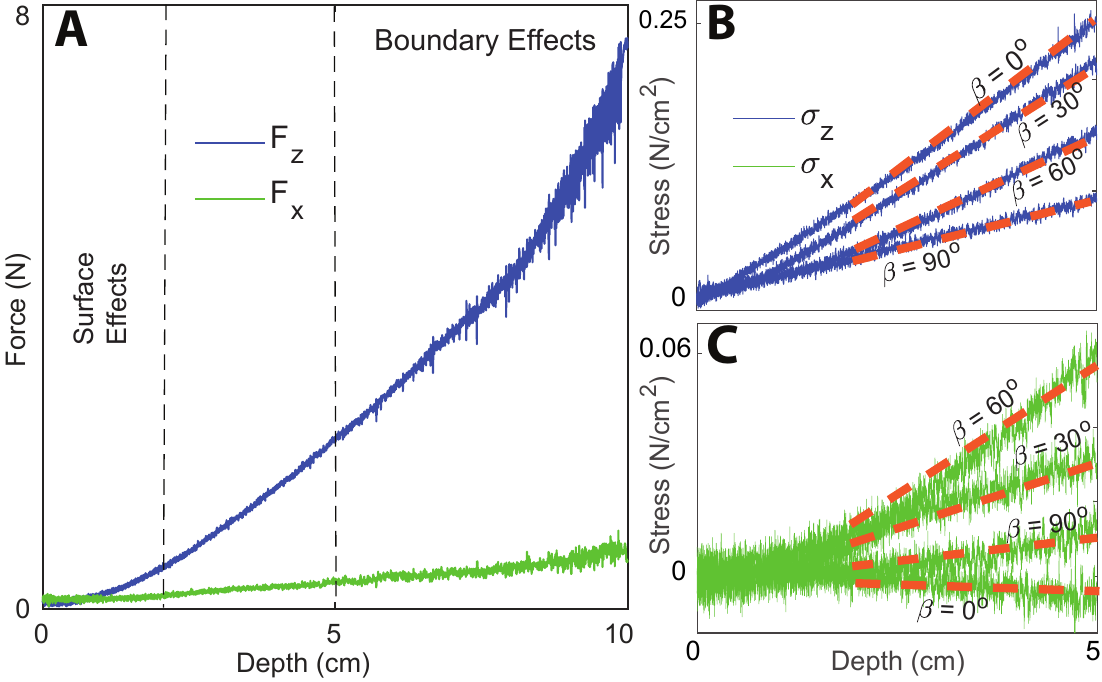}
    \caption{Force and stress plots during intrusions at various angles of attack. A) Force curves for z and x axes for $\gamma=90^\circ$ and $\beta = 30^\circ$. B) Stress curves (green and blue) for $\gamma=90^\circ$ and $\beta = [0^\circ, 30^\circ, 60^\circ, 90^\circ]$ and linear fits (dashed orange) to these curves.}
    \label{fig:RawForce}
\end{figure}

Forces in horizontal and vertical axes are shown in \refig{fig:RawForce}A. We observed that for all $\beta$ and $\gamma$ tested the forces were nearly proportional to depth $|z|$, similar to lithostatic pressure observed in non-cohesive granular media \cite{li2013terradynamics}. Using this region of depth gave us linear relations between depth and stress, dashed orange fits in \refig{fig:RawForce}B and \refig{fig:RawForce}C. 

\begin{figure}[!ht]
    \centering
    \includegraphics[width=1\columnwidth]{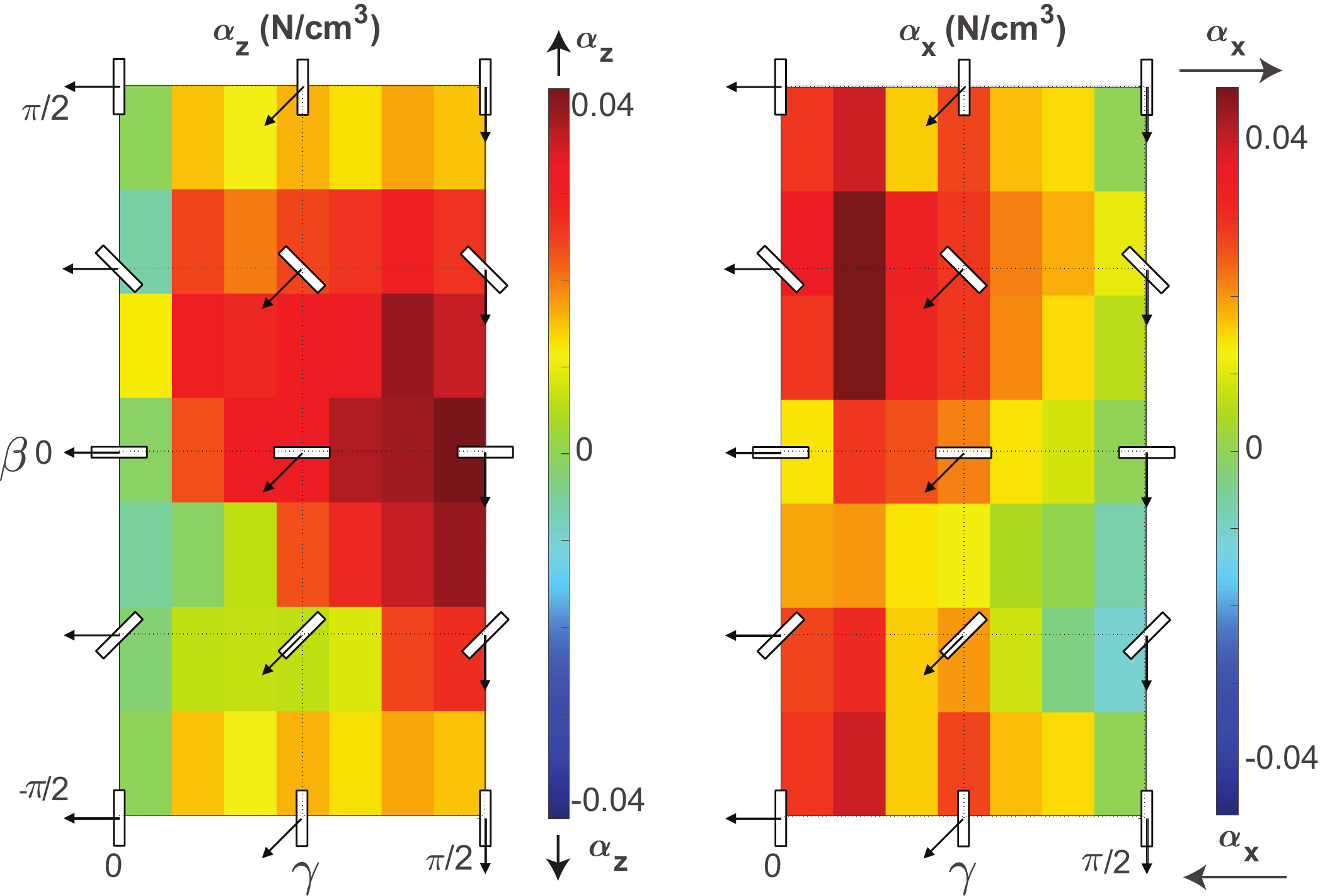}
    \caption{Stress-per-unit-depth, $\alpha_{z,x}$, relations for vertical and horizontal directions. The individual grids correspond to the slopes of the fitted lines for $0<\gamma<\pi/2$ and $-\pi/2<\beta<\pi/2$.}
    \label{fig:heatmap}
\end{figure}

These slopes are used to estimate the stiffness (stress per unit depth) of the substrate to the intruding plate, $\alpha_{z,x}$. The plate’s angle of intrusion $\gamma$ and attack $\beta$ are varied to obtain the vertical and horizontal stress per unit depth relations as illustrated in \refig{fig:heatmap}. Stresses in the powder, in both directions, were sensitive to both the angles of attack and intrusion, showing qualitative similarities to those observed in the non-cohesive granular heatmap. In general, the media resisted the plate’s motion across most values of $\beta$ and $\gamma$, as indicated by positive stress values in the $\sigma_{z,x}$ plots. However, for certain combinations of $\beta$ and $\gamma$, the material did not resist intrusion, which is reflected by negative values in the heatmap. For all $\gamma$ (except for $\gamma = \pi/2$), $\alpha_{z,x}$ were asymmetric to angle of attack values $\beta = 0$ and $\beta = \pi/2$. 

To interpolate intermediate stress values and compare cohesive and non-cohesive stress profiles we conducted a fitting approximation on powder stress per unit depth functions $\alpha_{z,x}$. Using the discrete Fourier Transform we obtained a fitting function in the form shown below, based on a similar approach in \cite{li2013terradynamics}. The functions can be approximated as:
\begin{equation}
    \alpha_z^{fit}(\beta,\gamma) = \sum_{m=-1}^1\sum_{n=0}^1[A_{m,n}cos{2\pi}(\frac{m\beta}{\pi}+\frac{n\gamma}{2\pi})+B_{m,n}sin{2\pi(\frac{m\beta}{\pi}+\frac{n\gamma}{2/pi})}]
\end{equation}
\begin{equation}
    \alpha_x^{fit}(\beta,\gamma) = \sum_{m=-1}^1\sum_{n=0}^1[C_{m,n}cos{2\pi}(\frac{m\beta}{\pi}+\frac{n\gamma}{2\pi})+D_{m,n}sin{2\pi(\frac{m\beta}{\pi}+\frac{n\gamma}{2/pi})}]
\end{equation}

\begin{figure}[!]
    \centering
    \includegraphics[width=.78\linewidth]{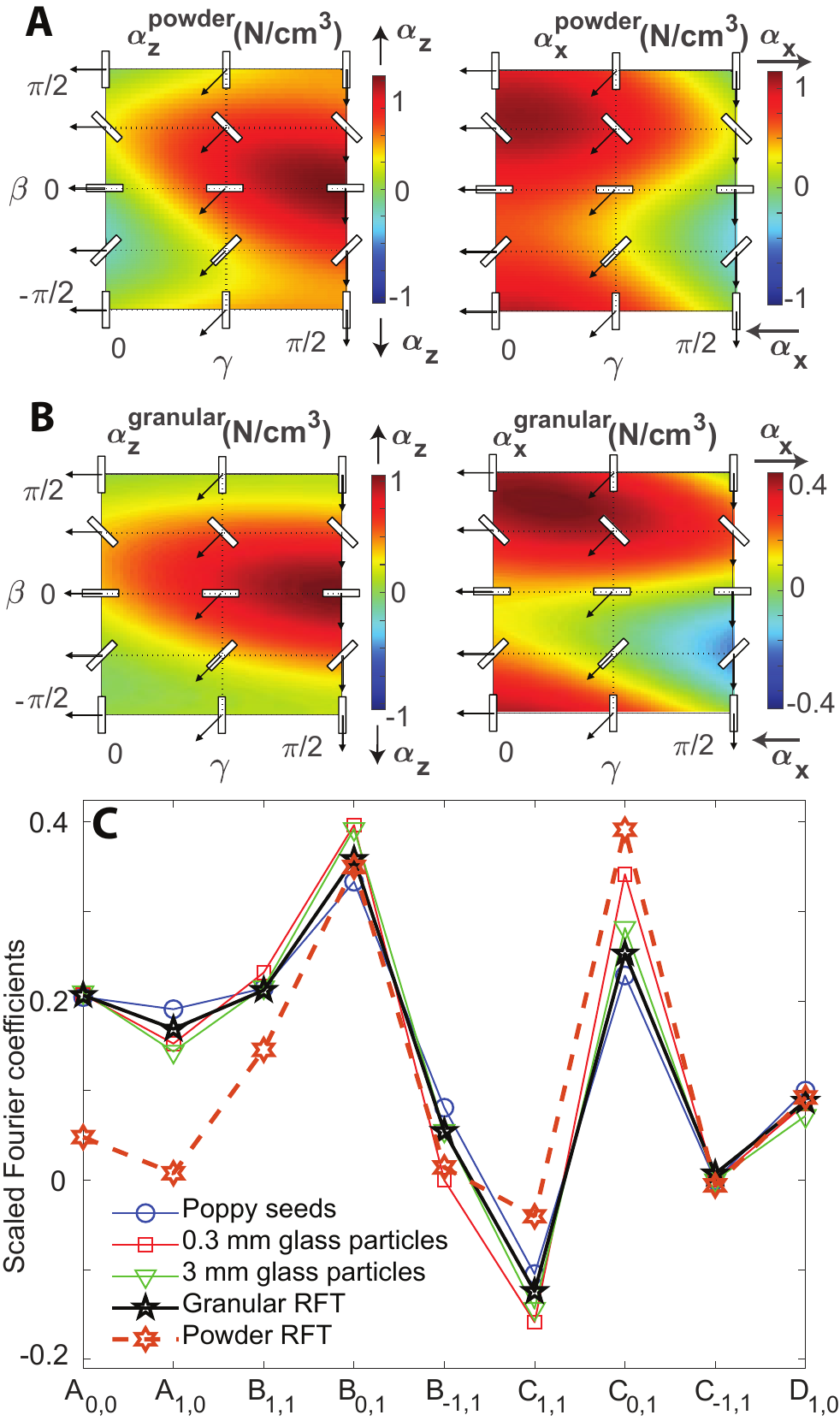}
    \caption{Plots of powder RFT (cohesive) (A) and granular RFT (non-cohesive) stress per unit depth profiles (B) and scaled Fourier coefficients of cohesive and non-cohesive granular media C) Scaled Fourier coefficients of various non-cohesive and cohesive media.}
    \label{fig:scale_hm}
\end{figure}

The powder stress profiles using the fitting function are plotted in \refig{fig:scale_hm}A, which we refer to as powder RFT (pRFT). We then adopted the granular stress per unit depth profiles from \cite{li2013terradynamics}, as shown in \refig{fig:scale_hm}B, representing granular media with particle sizes ranging from 0.3 mm to 3 mm for comparison purposes. Although qualitatively similar along both axes, we observed notable differences between the powder and granular heatmaps. Specifically, the material's resistance to penetration in the horizontal axis is significantly higher in the powder compared to the granular media. Additionally, the powder stresses exhibit larger resistance near $\beta = \pm \pi/2$ (the top and bottom regions of the heatmaps), unlike granular media, where resistance diminishes rapidly. In this dry powder regime, we attribute these differences to the cohesive nature of the material, which enables stronger inter-particle forces than non-cohesive granular media. The lack of attractive forces between particles leads to faster stress diminution. 

We then compared the scaled Fourier coefficients of the fitting functions in equations (1) and (2) for powder and granular RFT, along with various non-cohesive granular materials. \refig{fig:scale_hm}C shows the zeroth and first-order coefficients of these fitting functions. We observed that the coefficients for the powder exhibited similar trends to those of different non-cohesive granular materials. 

\subsection{Experimental Validation of pRFT}
\begin{figure}[!ht]
    \centering
    \includegraphics[width=.9\columnwidth]{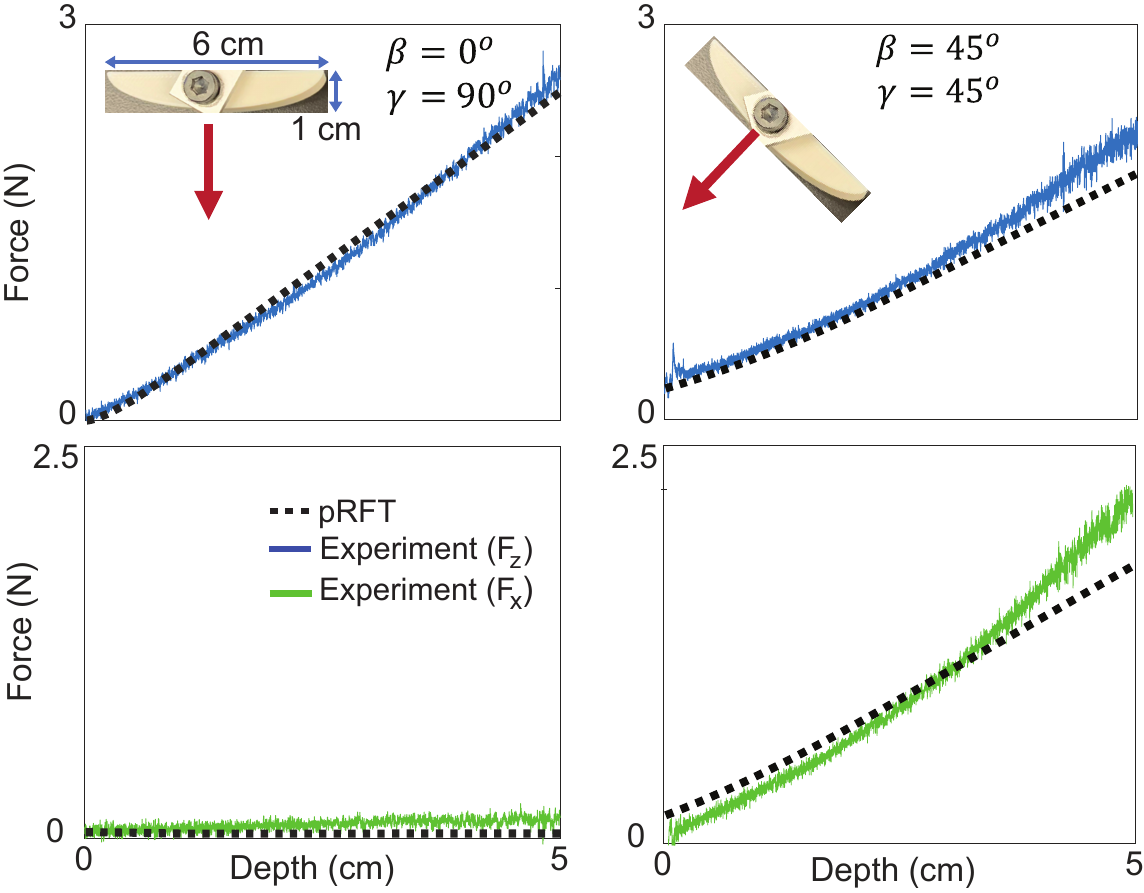}
    \caption{ Force curves obtained from powder intrusion experiments and powder RFT calculations for a non-trivially shaped rigid object. A comparison between the solid (experimental) and dashed (pRFT) curves shows that our model largely predicts the total forces acting on the body through linear superposition. } % XX MAKE ASPECT RATIO OF FIGS SQUARE SO EASIER TO SEE. NO NEED TO SQUASH EVERYTHING!
    \label{fig:footpad_model}
\end{figure}

As a next step, we tested if the forces acting on a complex intruder moving in powder could be estimated by applying linear superposition to the forces acting on small elements of the intruder. To this end, we conducted powder intrusion experiments with a curved geometry and measured its vertical ($F_z$) and horizontal forces ($F_x$) for two different $\beta$ and $\gamma$ scenarios (see \refig{fig:footpad_model}). We then performed powder RFT calculations by discretizing the geometry into 30 small plate elements and integrating the stresses along the curve's leading edge \cite{askari2016intrusion}. By plotting the measured (solid) and predicted (dashed) data in Fig. \ref{fig:footpad_model}, we demonstrated that the pRFT calculations closely matched both the horizontal and vertical forces for both intrusion configurations.

\subsection{Utilize pRFT to investigate lander footpad geometries}
In this section, we showcased the predictive abilities of our model by using it to discover geometries that enhanced resistance to intrusion into loosely consolidated cohesive powder. This is particularly relevant for the design of extraterrestrial landing gear, where landing on extremely weak surfaces poses a significant challenge to mission success. Specifically, Enceladus' loosely consolidated and cohesive surface consisting of ice plume deposits is a prime example of such a terrain. Here, we are interested in how various geometries generate different reaction forces through extensive pRFT calculations. In doing so, we aim to identify geometries that enable reduced sinkage during intrusion into weak surfaces.

\begin{figure}[!ht]
    \centering
    \includegraphics[width=1\columnwidth]{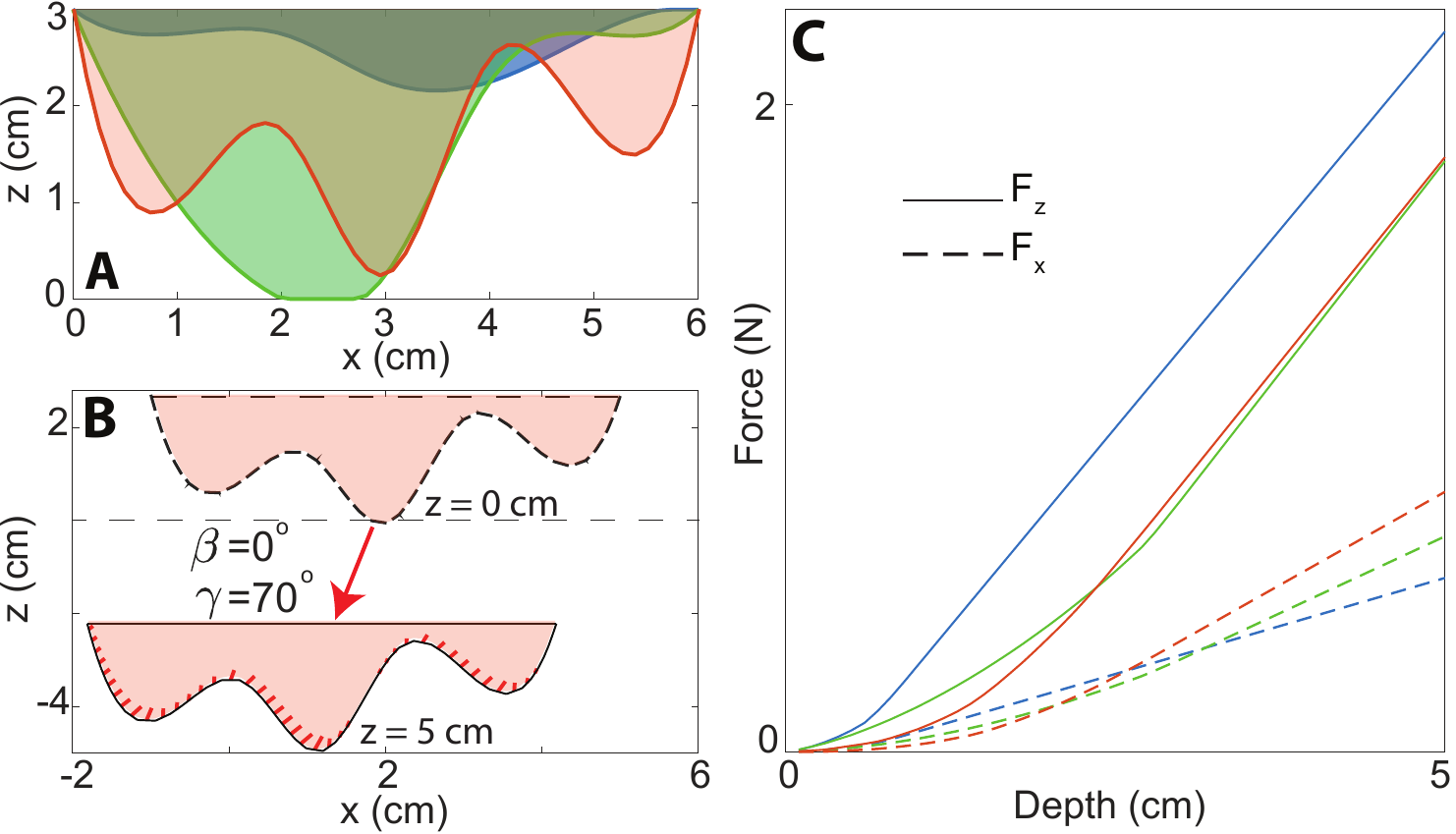}
    \caption{Illustrations and force profiles of footpad geometries. A) Sample footpad shapes illustrated as different colors generated via spline. B) Illustration of a sample intrusion to a depth of 5 cm for $\beta=0^\circ$ and $\gamma=70^\circ$. C) Forces of the footpad intrusion shown in \refig{fig:footpad_create}B for z (solid) and x-axess (dashed) using pRFT.}% XX MAKE ASPECT RATIO OF C HIGHER SO TOP AND BOTTOM LINES MATCH TOP AND BOTTOM OF A AND B} 
    \label{fig:footpad_create}
\end{figure}

We started by creating arbitrary footpad shapes by first meshing a 2D plane. The meshed region is 6 cm long (footpad length) and 3 cm deep (footpad depth) with grid sizes of 1 cm and 0.25 cm, respectively. Then, we fit splines along the intersections of these meshes giving us arbitrary shapes. \refig{fig:footpad_create}A illustrates three sample footpad shapes. Due to the principle of linear superposition, the forces generated by the small-scale footpad can reasonably scale to those produced by a larger footpad during ground contact in a landing event. Therefore, our investigation will provide valuable insights into the design of large-scale footpads for NASA JPL's landing missions.

Since our model only generates reaction forces as outputs, we intruded the geometries to a fixed depth to ensure a fair comparison among them. Starting from the surface, we intruded shapes to a depth of 5 cm for $\beta=0^\circ$ and $\gamma=70^\circ$ as illustrated in \refig{fig:footpad_create}B. \refig{fig:footpad_create}C illustrates the vertical (solid) and horizontal (dashed) force curves for three shapes, with the corresponding colors indicated in \refig{fig:footpad_create}A. The vertical forces were larger than the horizontal forces since all shapes move primarily in a vertical direction within the media. Although the red and green footpads have different shapes, they exhibited similar force trends, particularly in the vertical direction. The blue shape, having a minimal contact surface area among others, produced the largest vertical force and the smallest horizontal force. The large vertical forces observed in the blue shape can be attributed to its greater horizontal segments, generating increased resistance to vertical movement, as shown with the red regions in \refig{fig:heatmap}. In contrast, the sharp and the wavy segments in other shapes resulted in reduced vertical resistance, as observed in the yellow and green regions in \refig{fig:heatmap}. The reduced horizontal resistance of the blue footpad can be attributed to its narrow thickness, whereas the other shapes are thicker and have a larger resistance to horizontal motion. 

Next, we examined the resistive forces of all the arbitrary footpads we generated. Specifically, we calculated the vertical resistive forces at a given depth, as we focused on identifying shapes that generated maximal vertical forces. We conjecture that the greater the vertical resistive force a shape produces at a certain depth, the less sinkage the landing module will experience on a weakly consolidated surface. Thus, a shape that produces larger forces would result in less sinkage. We also explored shapes that produce minimal vertical force to gain a deeper understanding of the mechanisms behind force generation in the maximal-force shapes.

\begin{figure}[!ht]
    \centering
    \includegraphics[width=1\columnwidth]{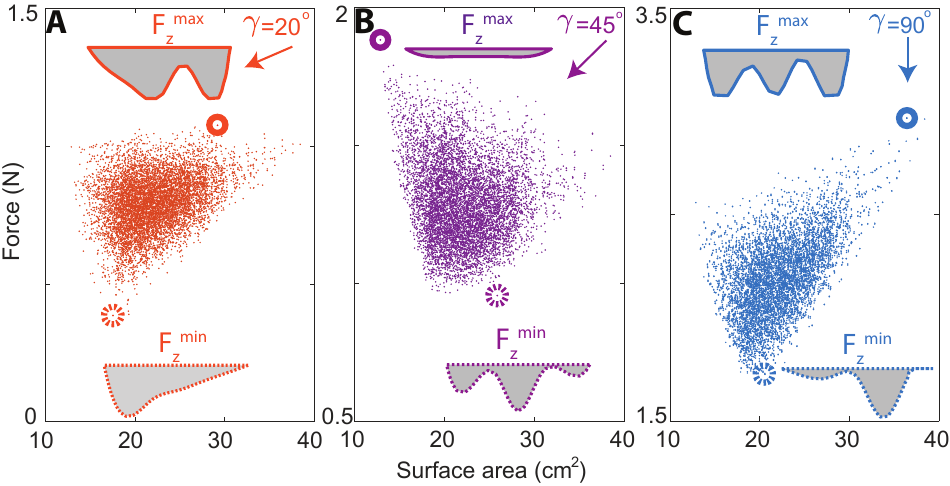}
    \caption{pRFT max force distributions of footpad geometries at an intrusion depth of 5 cm for three different angles of attack, illustrating the high variability in forces. Each point in \refig{fig:pRFT_Force}A-C, represents the vertical force of individual footpad geometries (a total of 7776) as a function of their respective surface areas at a depth of 5 cm for $\gamma=[20^\circ,~45^\circ,~ 90^\circ]$, shown from left to right, respectively.}
    \label{fig:pRFT_Force}
\end{figure}

To ensure a fair comparison among the footpads, we made several assumptions about their intrusion. First, we assumed that the footpads make contact on a level surface, with the footpad’s attack angle $\beta$ parallel to the surface. We then assumed that the footpad's orientation remained constant during intrusion. In other words, the RFT calculations assumed $\beta=0$ during the intrusion, as we cannot predict how much deflection the leg will experience during the actual planetary landing. Additionally, since the angle of intrusion during a planetary landing event is uncertain, we performed pRFT calculations for three $\gamma$ scenarios: perfect vertical intrusion ($\gamma=90^\circ$), diagonal intrusion ($\gamma=45^\circ$), and near horizontal intrusion ($\gamma=20^\circ$).

Our pRFT calculations showed that the range of vertical forces varied significantly, depending on both the footpad geometry and the angle of intrusion. This variation is more pronounced for near-horizontal intrusion (orange) than other intrusion scenarios (purple for diagonal and blue for perfect vertical intrusion). We highlighted the shapes that generate the maximal and minimal vertical forces by encircling their respective points via dashed and solid circles and visualized them at the top and bottom of \refig{fig:pRFT_Force}A-C. Examining the shapes that exhibit minimum forces across all intrusion scenarios revealed a common feature: they all have a large, sharp wedge-like form (shapes with dashed borders). The small vertical forces of wedge shapes can be attributed to their edges having higher angles of attack, which are associated with smaller stresses (yellow and green regions in \refig{fig:heatmap}). Additionally, in the near-horizontal intrusion scenario, the large wedge shape creates a shadowing effect \cite{suzuki2019study}, where forces are diminished in areas that lie in the immediate vicinity of another part of the intruder. In other words, the leading edge of the intruder (leftmost part), impedes the forces acting on the non-leading edge, which is the right part of the intruder.

In contrast, the shapes generating maximal forces did not share a common form across all intrusion scenarios (shapes with solid borders). However, curvy patterns that increase the contact surface area with the material tend to generate more force for $\gamma=20^\circ$ and $\gamma=90^\circ$. The increased surface area of the curvy footpad (solid blue and orange borders) likely contributes to this enhancement in force. Conversely, a flat surface with minimal surface area produced the largest vertical force in the diagonal intrusion and substantial forces in other intrusion scenarios (the leftmost point in all figures). This may be because most of the individual elements of flat shape penetrate to greater depths, generating more resistive force than more curvy shapes, whose curved edges penetrate the material gradually. Additionally, the curvy regions contribute to the shadowing effect, reducing the vertical forces.

\begin{figure}[!ht]
    \centering
    \includegraphics[width=1\columnwidth]{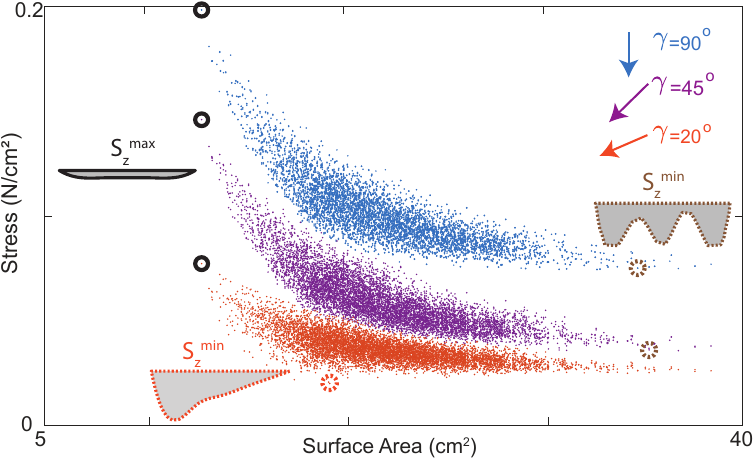}
    \caption{pRFT max stress distributions of footpad geometries at an intrusion depth of 5 cm for three different angles of intrusions for $\gamma=[20^\circ,~45^\circ,~ 90^\circ]$, shown as orange, purple, and blue dots respectively.}
    \label{fig:pRFT_stress}
\end{figure} %XX FONTS LOOK SQUEEZED ON Y AXIS AND LABELS TOO SMALL IN LITTLE INSETS

A key factor in footpad design is the mass of the structure. Ideally, the footpad geometry should be lightweight while providing high vertical resistance during landing. This requires evaluating the stress (force per unit area) generated by the footpad geometries, where the unit area can be considered as the unit mass of the footpad. Hence, a footpad geometry that generates higher stress would produce greater force per unit mass. To examine the stress profiles of the footpads, we plotted the stress as a function of their corresponding surface areas, as shown in \refig{fig:pRFT_stress}. The flat surface geometry (shape with a solid black border) produces the highest stresses in all intrusion cases. The smaller surface area of the flat-surfaced footpad contributes to the increased stress. In contrast, the curvy footpad (the shape with a dashed brown border) exhibited the minimal stress for $\gamma=45^\circ$ and $\gamma=90^\circ$ due to its larger surface area, while a large wedge-shaped footpad generated the minimal stress at $\gamma=20^\circ$. 

\subsection{Testing pRFT predictions of footpad geometries}
\begin{figure}[!ht]
    \centering
    \includegraphics[width=1\columnwidth]{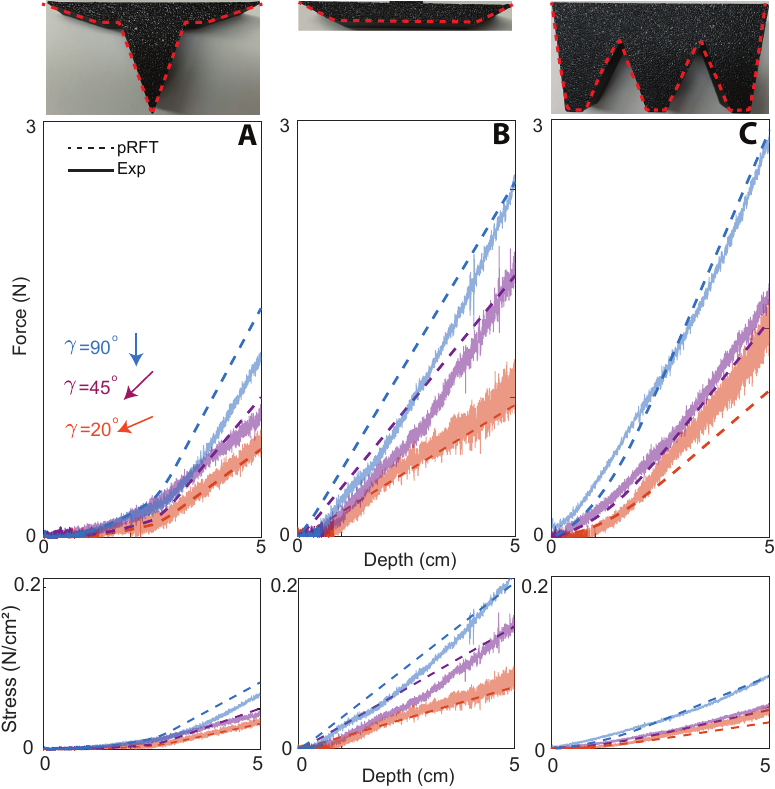}
    \caption{Force plots of intrusion experiments (solid) versus pRFT predictions (dashed) with three intruders under three different angles of intrusion. Insets represent the stress profiles of intruders. Powder intrusions of A) wedge B) flat and C) curvy geometries under near-horizontal (orange), diagonal (purple), and vertical (blue) angles of intrusion. }
    \label{fig:pRFT_validate}
\end{figure}

We validated the accuracy of our pRFT calculations by conducting intrusion experiments using some of the key footpads predicted by our model. We tested three footpad geometries—the wedge-shaped, flat-surfaced, and curvy shapes—as illustrated in \refig{fig:pRFT_validate}A-C. Each footpad is penetrated to a depth of 5 cm across three intrusion scenarios ($\gamma=[20^\circ,45^\circ,90^\circ]$). The plots in \refig{fig:pRFT_validate}A-C show the force-depth profiles for the tested shapes, with the insets depicting the corresponding stress-depth relationships. The solid curves depict the experimental trials for each $\gamma$ corresponding to the black shapes in the images, while the dashed curves illustrate the pRFT predictions, which correspond to the red dashed boundary lines in the images. 

Our pRFT predictions closely aligned with the experimental results; however, the footpads displayed varying degrees of mismatch under different intrusion conditions. Interestingly, the predictions for wedge shape (\refig{fig:pRFT_validate}A) showed the largest mismatch for $\gamma=90^\circ$, though pRFT was still able to qualitatively capture the force trend. The predictions for the flat surface (\refig{fig:pRFT_validate}A) were reasonably accurate; however, for $\gamma=45^\circ$ and $\gamma=90^\circ$, pRFT failed to capture the bowing trend observed in the force profiles. The forces induced by the curvy footpad (\refig{fig:pRFT_validate}A) are successfully captured by pRFT except for $\gamma=20^\circ$, where pRFT underpredicted the forces. This mismatch is thought to arise from the adjacency effects of the wedges \cite{pravin2021effect} and the high variability in the material's volume fraction during intrusion.

Among the shapes tested, the wedge footpad generated the lowest resistive force. The flat surface exhibited increased vertical resistance for all $\gamma$, whereas the curvy shape showed the highest resistance for $\gamma=90^\circ$, exceeding that of the flat surface. In non-vertical intrusion scenarios, the curvy shape's resistance was similar to that of the flat surface. Regarding the stress profiles, the flat surface outperformed the others, while the wedge and wavy shapes displayed lower and comparable stress distributions.

 %XX I DONT THINK YOU NEED TO CRAM THE FIGURE SO TIGHTLY. PUT THE IMAGES ON THE FIRST ROW, FORCE ON 2ND AND STRESS ON 3RD.

The results from our pRFT calculations and experiments indicate that if minimizing footpad mass is a priority, a flat surface provides the largest stress distribution across all intrusion scenarios tested. Conversely, if footpad mass is less constrained, a saw-tooth or curved design could be advantageous, as it increases surface area and generates more force, particularly under more vertical intrusion conditions. 

\section{Conclusion}
In this work, drawing on a hypothesis based on frictional plasticity theory and the success of granular RFT, we developed a model to predict forces on bodies moving within the dry cohesive powder. This involved calculating the substrate's stress per unit depth by measuring the stresses encountered by a small plate at various attack and intrusion angles and depths. By extending granular RFT, originally developed for non-cohesive media, to include cohesive powder, we verified that the linear superposition of elemental forces acting on an arbitrarily shaped intruder accurately predicts the total forces of the body moving in powder. We demonstrated that powder RFT stress profiles qualitatively align with granular RFT; however, distinct trends emerged between the cohesive and granular stress profiles. Specifically, we observed higher resistance to horizontal intrusions in cohesive powder compared to non-cohesive granular media. This effect resembled a more isotropic resistance to motion within the powder, which we attributed to the cohesion from interparticle forces in the powder substrate.

Motivated by the needs of NASA JPL's landing missions we tested our model's capabilities by calculating the powder resistive forces on systematically generated geometries via spline curves. Our goal was to identify geometries that enhance resistance to intrusion in loosely consolidated cohesive powder. By focusing on planetary lander footpads, we sought to determine which geometries minimize sinkage on weak surfaces, addressing the challenge of landing stability. This is particularly relevant for future missions to Enceladus and Europa, whose surfaces partly consist of loosely consolidated ice powder. Our model’s predictions, along with experimental data, indicated that a flat surface provides the highest force distribution per unit footpad mass across the intrusion scenarios tested. Alternatively, a curvy surface with an increased surface area and mass generated the largest force, particularly under more vertical intrusion conditions. More broadly, we posit that the pRFT calculations can be leveraged to design legged and wheeled systems capable of traversing on extremely weak and cohesive terrain.
\appendix

%\section{Appendix}
%\label{sec:appendix}

\section{Acknowledgment}
The research was supported by the Strategic University Research Partnerships (SURP) at the Jet Propulsion Laboratory, California Institute of Technology, under a contract with the National Aeronautics and Space Administration (80NM0018D0004). The authors would like to acknowledge Laura Treers for her valuable insights and suggestions and Juntao He for collecting the preliminary data.
 
\bibliographystyle{elsarticle-num} 
\bibliography{refs}
\end{document}